\documentclass[aps,prb,manuscript]{revtex4}
\usepackage{graphicx}
\usepackage{epsfig}
\usepackage{dcolumn}
\usepackage{bm}
\usepackage{setspace}

\begin{document}

\texttt{}

\title{Enhanced response of current-driven coupled quantum wells}
\author{Antonios Balassis }
\address{ Physics Department, Fordham University, 441 East Fordham Road,
       Bronx, NY 10458, USA}
       \author{Godfrey Gumbs}
\address{Department of Physics and Astronomy, Hunter College of the
City University of New York, 695 Park Avenue, New York, NY 10065, USA}

\date{\today}

\begin{abstract}

We have investigated the conditions necessary to achieve stronger
Cherenkov-like instability of plasma waves
leading to emission in the terahertz (THz) regime  for semiconductor
quantum wells (QWs). The surface response function is calculated for a bilayer
two-dimensional electron gas (2DEG) system in the presence of  a  periodic spatial modulation of the equilibrium electron density. The 2DEG layers are coupled to surface
plasmons arising from excitations of free carriers in the bulk region between the
layers. A current is passed through one of the layers and is characterized  by a
drift velocity $v_D$ for the driven  electric charge.  By means of a
surface response function formalism, the plasmon
dispersion equation is obtained as a function of frequency $\omega$, the
in-plane wave vector ${\bf q}_{\parallel}=(q_x,q_y)$ and reciprocal lattice vector $nG$
where $n=0,\pm1,\pm2,\cdots$ and $G=2\pi/d$ with $d$ denoting the period of
the density modulation. The dispersion equation, which yields  the  resonant frequencies, is solved numerically
in the complex $\omega$-plane for real wave vector ${\bf q}_{\parallel}$. It is ascertained
that the imaginary part of $\omega$ is enhanced with decreasing $d$, and with increasing
the doping density of the free carriers in the bulk medium for fixed period of the spatial modulation.

\end{abstract}


\maketitle

\section{Introduction}
\label{sec1}

The abundance and availability of information in the literature
today is both a challenge and a blessing to physicists interested
in studying sources of terahertz (THz) radiation.  One THz (= $10^{12}$ Hz)
encompasses frequencies invisible to the naked eye in the electromagnetic spectrum,
lying between microwave and infrared. Gone are the days when experimentalists
relied heavily on "intuition" in their quest for stronger and more uniform sources
of radiation. Instead, today's experimentalists  are surrounded by a flood  of
published material yearly. Harvesting this abundance of knowledge and using it
to inform and improve this  area of research  presents a challenge for
researchers. It is imperative for experimentalists to review,
critique and transfer scientific findings to effective quality outcomes.
One active field being pursued is that of THz quantum cascade lasers (QCLs).
QCLs were invented in 1994 \cite{1}, are semiconductor lasers obtained by
epitaxially growing a sequence of layers of different semiconductors. As a result,
electrons in the conduction band are subjected to an artificial, periodic,
one dimensional potential that varies on the nanometer scale.
So far, the focus has been on the design, growth, fabrication, and testing of
high power THz QCLs emitting across a broad frequency range. The work reported
up to this point covers several ambitious projects from ultra-long wavelength emission,
phase/mode-locking, multiple color generation, photonic crystal structures, and
improved laser performance with respect to both maximum temperature operation and peak output power.

Far-infrared (FIR) radiation has been used to excite
plasmon modes resonantly in a quantum-well transistor at frequencies between
0.5 and 1.0$\times10^{12}\ s^{-1}$. A split grating gate design that allows localized
pinch-off of the transistor channel has been found to significantly enhance FIR response
and to allow  electrical tuning of the plasmon resonance.\cite{2,3} As a matter of
fact, the role played by plasma excitations in the THz
response of low-dimensional microstructured semiconductors
has been investigated considerably in recent times.\cite{4,5,6,7,8,9,10,11,12,13}
This paper discusses a specific set of simulations which bolster
earlier work which has already been published. We carry out a comprehensive study and
report our results for investigating plasma instabilities in a pair
of Coulomb coupled two-dimensional electron gas (2DEG) systems in which
inter-layer hopping between layers is not included.\cite{12,13}

\section{General Formulation of the Problem}
\label{sec2}

Our system consists of a pair of parallel two-dimensional electron gas layers embedded
 in the half-space $z\geq0$.  The space between the layers is filled by a dielectric
 material with dielectric constant $\epsilon(\omega)$. In Fig.~1
  we show a schematic   of the bilayer structure which we do numerical
  calculations for. However, in our formulation of the problem, we consider an arbitrary
  number of parallel 2DEG layers. We assume that our multi-layered structure interacts
  with an external time dependent electrostatic potential $\phi_{\rm ext}({\bf r},t)$.

We denote the  electric field due
to an external electrostatic potential by ${\bf E}_{\rm
ext}=-\bm{\nabla}\phi_{\rm ext}$. Let the layered system  occupy
the region $z>0$. When we periodically modulate the equilibrium electron density, the surface response
function will be modified from its value in the absence of the modulation.\,\cite{12,13}
We assume that the periodic density modulation is along the $x$-direction and
that the modulated sheet density can be described by $n_{\rm
2D}(x)=\sum\limits_{n=-\infty}^{\infty}\rho_n\exp(inGx)$, where the
real-value $\rho_n$ is the $n$th Fourier component of $n_{\rm
2D}(x)$, $G=2\pi/d$ is the reciprocal lattice vector and $d$ is the
period of the modulation.
Since $\nabla^2\phi_{\rm ext}=0$ for $z>0$, the electrostatic potential in this region and in the
vicinity of the surface  can be written as a superposition of plane
waves. This follows from the fact that in the region outside the surface,
where there is no charge present, the external potential
$\phi_{\rm ext}$ satisfies $\nabla^2\phi_{\rm ext}=0$.
Since the system is translationally invariant in the $y$-direction but periodic
parallel to the $x$-axis, we have

\begin{equation}
\phi_{\rm ext}({\bf r};\omega)=\sum_{n=-\infty}^\infty \
\int \frac{d^2{\bf q}_\parallel}{(2\pi)^2}\
 e^{iq_{x,n}x+iq_yy}\tilde{\phi}_{\rm ext}(q_{n},q_y,z;\omega)\ ,
\label{A1}
\end{equation}
where $q_{x,n}=q_x+nG$, $q_n=\sqrt{q_{x,n}^2+q_y^2}$, and $\tilde{\phi}_{\rm ext}(q_n,q_y,z;\omega)$ satisfies $q_n^2\tilde{\phi}_{\rm ext}=d^2\tilde{\phi}_{\rm ext}/dz^2$. Consequently, the general solution just outside   ($z\stackrel{<}{\sim}0$) or just within
($\stackrel{>}{\sim}$)   the bi-layer system  is given by

\begin{displaymath}
\phi_{\rm ext}({\bf r};\omega)=\sum_{n=-\infty}^\infty \
\int \frac{d^2{\bf q}_\parallel}{(2\pi)^2}\ \ \left[ \Gamma_+(q_n;\omega)
e^{  q_nz} +\Gamma_-(q_n;\omega)
e^{\pm q_nz}   \right]\ e^{iq_{x,n}x+iq_yy}
\end{displaymath}
where the upper (lower) sign in the exponential factor
$e^{ \pm q_nz} $   may be chosen when $z<0$ or $z>0$ for the solution of
Poisson's equation.  The pair of coefficients  $\Gamma_\pm(q_n;\omega)$   should
be chosen to have different values above and below the plane at $z=0$ and are
determined by the boundary conditions.
on the electrostatic potential and electric field
Also,  ${\bf q}_{\parallel}=(q_x,q_y)$ is an in-plane 2D wave vector.
The induced potential is obtained by solving
$\nabla^2\phi_{\rm ind}=-\rho_{\rm ind}/\epsilon_0$. This may be rewritten as

\begin{eqnarray}
\phi_{\rm ind}({\bf r}^\prime;\omega)&=&\int d{\bf r}^\prime \
v({\bf r},{\bf r}^\prime)\rho_{\rm ind}({\bf r}^\prime;\omega)
\nonumber\\
&=& \int d{\bf r}^\prime\ d{\bf r}^{\prime\prime} \ v({\bf r},{\bf r}^\prime)
\chi^{(0)}({\bf r}^\prime ,\ {\bf r}^{\prime\prime};\omega)\
\phi_{\rm ext}({\bf r}^{\prime\prime};\omega)\ ,
\label{A3}
\end{eqnarray}
where $v({\bf r},{\bf r}^\prime)$ is the Coulomb interaction potential
between the electrons in the 2D system,
and  $\chi^{(0)}({\bf r}^\prime ,\ {\bf r}^{\prime\prime};\omega)$ is the nonlocal,
frequency-dependent density-density response function. By Fourier transforming
the response function and the external potential in the variables $x$ and $y$, we
obtain after a straightforward calculation the result

\begin{equation}
\phi_{\rm ind}({\bf r};\,\omega)=-\sum_{n=-\infty}^\infty\,
\int \frac{d^2{\bf q}_\parallel}{(2\pi)^2}\
g_n({\bf q}_{\parallel};\,\omega)\,e^{q_nz}\,e^{iq_{x,n}x+iq_yy}
\ , \label{A4}
\end{equation}
where the Fourier component of the surface response function is
defined by

\begin{equation}
g_n({\bf q}_{\parallel};\,\omega)=-
\int dz^\prime \int dz^{\prime\prime}\ e^{-q_nz^\prime}\
\chi^{(0)}(z^\prime,z^{\prime\prime};q_{x,n},q_y;\omega)\phi_{\rm ext}(q_{n},q_y,z^{\prime\prime})\ .
\label{A5}
\end{equation}
Combining the results in Eqs.\ (\ref{A2}) and (\ref{A4}), we obtain
the total potential, which is the sum of the external and induced
potential outside the layered system

\begin{equation}
\phi({\bf r};\,\omega)=\sum_{n=-\infty}^\infty\, \int \frac{d^2{\bf q}_\parallel}{(2\pi)^2}
\left[e^{-q_nz}-g_n({\bf
q}_{\parallel};\,\omega)\,e^{q_nz}\right]\,e^{iq_{x,n}x+iq_yy}\ \ \ \
\mbox{for $z\stackrel{<}{\sim}0$}\ , \label{efst}
\end{equation}
where $q_n$, $q_{x,n}$ are defined above.  We must now calculate the Fourier components of the
surface response function $g_n({\bf q}_{\parallel};\,\omega)$ for a layered
2DEG. There is a dielectric medium  with dielectric constant
$\epsilon$ filling the space between adjacent layers, except for a
very thin vacuum region right next to the 2D layer. The
electrostatic potential  in the vacuum region at $z=z_\ell$ between
layers at $z=z_\ell$ and $z=z_{\ell+1}$ is given by $\phi_\ell({\bf
r},\,t)=e^{-i\omega t}\,\phi_\ell({\bf r},\,\omega)$, where

\begin{equation}
\phi_\ell({\bf r};\,\omega)=\sum_{n=-\infty}^\infty\,
\int \frac{d^2{\bf q}_\parallel}{(2\pi)^2} \left[
A_\ell^{(n)}e^{-q_{n}(z-z_\ell)}+B_\ell^{(n)}e^{q_{n}
(z-z_\ell)}\right]\,e^{iq_{x,n}x+iq_yy}\  . \label{eqgg2+}
\end{equation}
Also, at the last interface we take

\begin{equation}
\phi_{L+1}({\bf
r};\,\omega)=\sum_{n=-\infty}^\infty\, \int \frac{d^2{\bf q}_\parallel}{(2\pi)^2}\
e^{iq_{x,n}x+iq_yy}\,t_{L+1}\,
e^{-q_{n}(z-z_L)}\ \ \ \ \mbox{for $z\geq z_{L+1}$}\
\label{eqgg8+}
\end{equation}
where the coefficient $t_{L+1}^{(n)}$ must be determined
from the matching boundary conditions
like $A_l^{(n)}$ and $B_l^{(n)}$  However, we only choose the exponentially
decaying term to ensure that the solution for the potential does  not
blow up.

Both the potential $\phi$ and its derivative $\epsilon d\phi/dz$
must be continuous at the vacuum-dielectric medium interface.
However, the electric field is discontinuous across the 2D charged
layer on which the induced electron sheet density is

\begin{equation}
\sigma_\ell^{(n)}(q_n,\,\omega)=e^2\chi_{\ell}^{(0)}(q_n,\,\omega)\left(A_\ell^{(n)}+B_\ell^{(n)}\right)\
, \label{eqgg3}
\end{equation}
where $\chi_{\ell}^{(0)}(q_n,\,\omega)$ is a Fourier component of the single-particle
density-density response function of the $\ell$-th layer. We note that we employ no other parameter
to describe the gated grating period.  As a consequence, we only have to
 specify the grating potential in our numerical calculations
 by the period $d$.  After some straightforward algebra, we obtain

\begin{equation}
\left[\begin{array}{cc}
A_{\ell+1}^{(n)} \\
B_{\ell+1}^{(n)}
\end{array}   \right]
=\tensor T^{(n)}(\alpha_{\ell})  \left[\begin{array}{cc}
A_{\ell}^{(n)} \\
B_{\ell}^{(n)}
\end{array}   \right]={\cal M}_\ell^{(n)}  \left[\begin{array}{cc}
1 \\
-g(q_n,\,\omega)
\end{array}   \right]
 \ ,
\label{eqgg4}
\end{equation}
where $g(q_n,\,\omega)\equiv g_n({\bf q}_{\parallel};\,\omega)$ depends
only on the total wave number $q_n=\sqrt{q^2_{x,n}+q_y^2}$, ${\cal
M}_\ell^{(n)}= \tensor T^{(n)}(\alpha_\ell)\bigotimes
\cdots\bigotimes\tensor{T}{(n)}(\alpha_1)\bigotimes \tensor
T^{(n)}(\alpha_0)$, and

\begin{eqnarray}
\left[\tensor T^{(n)}(\alpha_\ell) \right]_{11} &=&\left[
(1+\epsilon)(\epsilon+1+2\epsilon\alpha_\ell)\,e^{-q_n a}+
(1-\epsilon)(\epsilon-1-2\epsilon\alpha_\ell)\,e^{q_na} \right]/4\epsilon
\nonumber\\
\left[ \tensor T^{(n)}(\alpha_\ell) \right]_{12}&=&\left[
(1+\epsilon)(\epsilon-1+2\epsilon\alpha_\ell)\,e^{-q_n a}+
(1-\epsilon)(\epsilon+1-2\epsilon\alpha_\ell)\,e^{q_n a}
\right]/4\epsilon
\nonumber\\
\left[\tensor T^{(n)}(\alpha_\ell) \right]_{21}&=&\left[
(1-\epsilon)(\epsilon+1+2\epsilon\alpha_\ell)\,e^{-q_n a}+
(1+\epsilon)(\epsilon-1-2\epsilon\alpha_\ell)\,e^{q_n a}
\right]/4\epsilon
\nonumber\\
\left[\tensor T^{(n)}(\alpha_\ell) \right]_{22}&=&\left[
(1-\epsilon)(\epsilon-1+2\epsilon\alpha_\ell)\,e^{-q_n a}+
(1+\epsilon)(\epsilon+1-2\epsilon\alpha_\ell)\,e^{q_n a}
\right]/4\epsilon\ . \label{eqgg5}
\end{eqnarray}
In Eq.\,(\ref{eqgg5}),
$\alpha_{\ell}(q_n,\,\omega)=(e^2/2\epsilon_0\epsilon)\,\chi_{\ell}^{(0)}(q_n,\,\omega)$
is the polarization function for each 2D charged layer.  When we
equate the electrostatic potential just inside the material at the
last layer to the electrostatic potential outside, we obtain from
Eq.\,(\ref{eqgg5}), after solving for the surface response function

\begin{equation}
g(q_n,\,\omega)=\frac{[1-\epsilon-2\epsilon\alpha_{L+1}(q_n,\,\omega)]{\cal
M}_{11}^{(n)}-[1+\epsilon+2\epsilon\alpha_{L+1}(q_n,\,\omega)]{\cal
M}_{21}^{(n)}} {[1-\epsilon-2\epsilon\alpha_{L+1}(q_n,\,\omega)]{\cal
M}_{12}^{(n)}- [1+\epsilon+2\epsilon\alpha_{L+1}(q_n,\,\omega)]{\cal
M}_{22}^{(n)}}\ , \label{eqgg6}
\end{equation}
where we obtain the elements of the $(2\times 2)$-matrix $\tensor
{\cal M}^{(n)}$ by  evaluating the product of $L$ transfer matrices
$\tensor T^{(n)}$ whose elements are defined in Eq.\,(\ref{eqgg5})
for a structure containing $L+1$ layers. Clearly, in this formalism,
the 2D charged layers are coupled through the Coulomb interaction
and we assume that {\em there is no interlayer electron hopping\/}
between layers. The surface response function in Eq.\,(\ref{eqgg6})
is a useful tool for calculating the normal mode spectrum of plasmon
excitations for a finite number of layers and can also be employed
to investigate the role played by layer separation on the loss
function ${\rm Im}\,[g(q_n,\,\omega)]$.\,\cite{radovic} The plasmon
dispersion is obtained by setting the denominator in
Eq.\,(\ref{eqgg6}) equal to zero. That is, the plasmon resonances
can occur for all values of $n$ and are determined by the angle the
in-plane polarization of the incident electromagnetic field makes
with the $x$-axis.

If there is a an electric field applied on the $\ell$-th layer creating a current and a drift velocity $\bf{v_D}$ , then there is a Doppler shift on the angular frequency of the density-density response function of this layer and we have to replace $\omega$ by $\omega-{\bf q}_n\cdot{\bf v_D}$, where ${\bf q}_n=(q_{x,n},q_y)$.

In the special case when there is a single
layer,  we set ${\cal M}_{11}^{(n)}={\cal M}_{22}^{(n)}=1$ and
${\cal M}_{12}^{(n)}={\cal M}_{21}^{(n)}=0$ in Eq.\,(\ref{eqgg6})
which becomes\,\cite{Eguiluz_Quinn,Persson}

\begin{equation}
g_{\rm single}(q_n,\,\omega)=1-\frac{2}{1+\epsilon +2
\epsilon\alpha(q_n,\,\omega)}\ . \label{eqgg6+}
\end{equation}
The surface response function for a semi-infinite slab of dielectric
medium can then be deduced from this result by setting the
polarization function $\alpha(q_n,\,\omega)$ equal to zero.

\section{Numerical Results and Discussion}
\label{sec3}

In this section, we present the results of our numerical calculations
of the real and imaginary parts of the plasmon frequency for a bilayer 2DEG.
Since inter-layer hopping is not included in  our model,
the layer separation must be chosen sufficiently large to satisfy this condition.
This requires that the inter-layer separation is much larger than
the lattice constant of the host material, which we assume to be GaAs/AlGaAs.
In our calculations, we chose the separation between the two parallel layers
of 2DEG to be $a=100${\AA}. The electron effective mass is chosen as $0.067\ m_e$,
appropriate for GaAs, where $m_e$ is the free-electron mass, the electron density
of each layer is taken as $n_{\rm 2D}=2\times10^{11}\; \textrm{cm}^{-2}$ which is
typical for a 2DEG. The corresponding
Fermi energy and Fermi frequency are $E_F=7.14\; \textrm{meV}$ and
$\omega_F=10.85\; \textrm{s}^{-1}$, respectively. We assume that the electrons
in one layer have a drift velocity $v_d=2.5\ v_F$. The region between the two
layers is filled with a dielectric material. The dielectric function of this material
is approximated by the local frequency-dependent form, i.e.,
 $\epsilon(\omega)=1-({\omega_p/\omega})^2$, where
$\omega_p=\left(4 \pi n_{\rm 3D}e^2/\epsilon_s \epsilon_b m^{\ast}\right)^{1/2}$
is the 3D plasma frequency of the dielectric material with background
dielectric constant $\epsilon_b$ and $\epsilon_s=4\pi\epsilon_0\epsilon_b$.
In our numerical calculations, the value that we used was $\omega_p=0.92\, \omega_F$.
The whole system is embedded in a medium with dielectric constant $\epsilon_b$=13.1.\

We justified using values for the  drift velocity greater than  the electron Fermi velocity
by a factor of 2.5  by noting that we are in a region of Cherenkov-like instability
of plasma waves  where, the drift velocity may be greater than the phase velocity.
Moreover, we have employed the
force-balance equation \cite{ting} as an alternative to the Boltzmann equation
to calculate the drift velocity as a function of applied electric
field and temperature in studying  on nonlinear electron transport in impurity-limited quantum
wires.   This system is less computationally intensive but,
nevertheless, gives a reasonable estimate of the drift
velocity as a function of strong electric field and as well as at various temperatures.
For example, when the ratio of the impurity to electron concentration
is $0.01$ and the temperature $T\approx 100$ K,  the drift
velocity can vary as $1.0 \times 10^7\ cm/s \leq v_d\leq 3.0 \times 10^7\ cm/s$
for an electric field which varies over the range
$0.5\ V/m\leq E\leq 2.4\ V/m$.  For the 2DEG density and the
electron effective mass chosen, the value of the Fermi velocity
is $v_F=2.09\times 10^7$ cm/s. Therefore, $v_D$ which we chose is reasonable in the frame of the force-balance formalism.

We first examined the effect which we obtain  on the complex plasmon
solutions by varying $n_{\rm 3D}$.  In Fig.~2,
we plot the real (black dots) and imaginary (red dots) parts
of the complex solutions $\omega$ (in units of $\omega_F$) of the plasmon dispersion
equation in terms of the in-plane wave vector $q_{\parallel}$  (in units of $k_F$) for the
parameters given above and for $\omega_p=0.92\omega_F$. No grating was  applied in these calculations.
Only those solutions with a {\em nonzero imaginary part\/} are shown in these plots.
Then, in Fig.~3, we increased $\omega_p$ to twice its value,
i.e., we chose $\omega_p=1.84\,\omega_F$ which corresponds to an increase
of $n_{\rm 3D}$ by a factor of 4. All the other parameters were kept the same
as in Fig.~2.  We can see, by comparing Fig.~2 with
Fig.~3, that $\Im m\ \omega$ in Fig.~3 has its local
 maximum at a higher value and exists over a wider range of wave vectors $q_{\parallel}$.
 Specifically, in Fig.~2, the imaginary part of $\omega$ has a
  maximum at $q_{\parallel}\approx 0.38\,k_F$ for which $\Im m\ \omega \approx 0.37\omega_F$.
  In Fig.~3, we see that the maximum
is located at $q_{\parallel}\approx0.62\,k_F$ where  $\Im m\ \omega \approx 0.57\omega_F$.
The increase in the value of
$\Im m\ \omega$ is by more than 50\%. Therefore, the electronic collective
excitations become more unstable, since their lifetime is inversely proportional to the imaginary part of the frequency. This enhancement of $\Im m\ \omega$
is reasonable since, in this case, an increase in $n_{\rm 3D}$ results in stronger
interactions and stronger coupling between the two layers. We found three
plasmon excitation modes, but in the figures we present two of them
which have nonzero imaginary frequency. Both of them are acoustic, while
the third mode which is not shown is optical.
The mode which is the most unstable in both figures is the least energetic
acoustic one. The mode just above this one, has a non-zero
imaginary part too which is much smaller than the lowest one.
The lowest plasmon mode bifurcates at $q_{\parallel}\approx 0.7 k_F$.
However, only the higher branch in this bifurcation has nonzero imaginary part and
only this is shown. Compared with the model in Ref.\ [\onlinecite{B1}], we
include a spatial separation between the 2DEG layers and passed a current
through one of them. This physical separation leads to the existence of
the appearance of two unstable modes.  The optical mode arises from the surface
plasmon mode and is robust against excitations. This is  unlike the acoustic
modes which come about from the in-plane charge density fluctuations for which
the screening properties are much different than the bulk.

We next turn to the results of our investigation for a bilayer with a spatially modulated
electron density. As  described above in Sec.\ \ref{sec2}, the modulation
is applied along the $x$-direction. The period was taken to be
$d=5\; \mu \textrm{m}$ which corresponds to a reciprocal lattice vector
$G=2\pi/d\approx1.26\times10^6\ \textrm{m}^{-1}$. This chosen value
for $d$ is a typical   experimental  characteristic for
metal gratings. The other parameters were
chosen the same as in Fig.~2. We  solved the dispersion equation
$g(q_n,\omega)=\infty$, where $g$ is the surface response function of the bilayer, $q_n=\left[(q_x+nG)^2+q_y^2)\right]^{1/2}$ is an in-plane wave vector and $n$
is an integer. For $n=0$, we have the case of the unmodulated bilayer which we mentioned
above in Figs.~2 and 3.  We are interested in the effect which the modulation
has, i.e., for the case $n=\pm1,\pm2,\ldots$. In
Fig.~4 (a), we plot $\Im m\ \omega$  as a  function of $q_{x}$
for three different values of $n=0,5\ \textrm{and}\ 10$ when $q_y=0$.
We see that as $n$ assumes larger positive values, the corresponding
$\Im m\; \omega$ curves are shifted to higher values of $q_x$ with negligible change
in their local maximum value. In order to better understand the meaning
of this shift, we examine a fixed $q_x=0.6\ k_F$. We then see that the curve
that corresponds to $n=10$ (blue dots) has a much larger $\Im m\; \omega$
than the curves for $n=0$ (black dots) and $n=5$ (red dots). Therefore,
the presence of the modulation leads to a shifting which in turn produces
 an amplification of the instability for values of $q_x\ge0.43\ k_F$ where
 the set of three curves corresponding to the acoustic branches
 intersect. However, for $q_x\le0.43\ k_F$, we have the
 opposite effect leading to an attenuation. For negative values of $n$,
  we see from Fig.~4 (b),  that the picture is more complicated
  since we have in addition to the $q_x$-shift, a change in the shape of
  the curve for high values of $n$. As we can see though for low values of $n$,
   the result is the same as before, i.e., there are intervals of $q_x$
    where we have amplification and intervals where we have attenuation.
    This is a result of the constructive or destructive mixing of the
     transmitted and reflected electromagnetic fields.  This is determined
     by the wavelength of the applied field and, as a result, leads to some
     range of wave vectors when there is  attenuation and another range
     where there is amplification.
Comparing now Fig.~2  to Figs.~4 (a) and
    4 (b), we can conclude that the presence of the modulation can
    give rise  to an  enhancement in the value of the
    $\Im m\; \omega$ curves as a result of the stronger mixing
    of the incident and reflected electromagnetic waves.

The effect on the imaginary part of the plasmon frequency $\omega$ when
we  vary the period $d$ of the modulation is shown in Figs.~5 (a)
and 5 (b) for fixed $n=2$ and $n=-2$, respectively. In Fig.~5 (a),
we see that  as the period of the modulation increases from $d=0.5\ \mu$m (black dots)
to  $d=500\ \mu$m (green dots), there are again intervals of $q_x$ where we
have enhancement and intervals where we have attenuation. In addition,
the curves approach a limiting value as the period $d$ increases. The behavior is
given approximately by the curve corresponding to $d=500\ \mu$m. This can be
explained by calculating the reciprocal lattice vector $G=2\pi/d$ which is
 negligibly small as $d$ becomes large. Large values of $d$ correspond to the
 case that there is no modulation, or equivalently to the case that $n=0$.
 Therefore, the limiting curve represents the response of the system when there
 is no spatial modulation from a grating. For $n=-2$, we see from Fig.~5 (b)
 that the picture is almost the same with the only difference
  being that the curve corresponding to $d=0.5\ \mu$m is deformed from its
  parabolic shape. In both figures, there is an increase in the height of the
  curves when we increase the period from $d=0.5\ \mu$m to higher value. We emphasize that
  the significance of ``$n$" is that it denotes the Bragg order
numbers in the presence of the modulation. The optical polarization and the
interference of a pair of  optical-polarization waves are is, of course, determined
by the value of $n$. This is why opposite values of $n$ can lead to different results
since the polarization depends on the magnitude and sign of the Bragg number.

\section{Concluding Remarks}
\label{sec4}
 In this paper, we
 used a linear response function  approach to investigate the plasmon instability for a bilayer
 system. In one of the 2DEG layers,   a steady current is passed with drift velocity
 ${\bf v}_D$ which induces a Doppler shift ${\bf q}\cdot{\bf v}_D$ in the
 frequency-dependent polarization function.  Here, we solved the plasmon dispersion
 relation in the complex frequency plane, following the work of Bakshi,
 et al.\cite{B1,B2,B3} We extended their work by employing a model
which couples  the 2DEG layers to a bulk plasma which could be doped with
free carriers and also by including the effect due to electron density
modulation.  The electrons in each
2D layer interact with the bulk plasma sandwiched between them.  The crucial
finding here is that there are {\em two} plasmon branches whose excitation
frequency  has nonzero imaginary part. They are both acoustic-like in nature.  The lower-frequency mode is more unstable than the one  with higher frequency. The  optical mode, which is not shown in the figures, is a result of the hybridization between the surface plasmon and the 2DEG plasmon. We have also demonstrated how the instability of
these modes is affected by varying the period of the modulation and the reciprocal
lattice vector which couples the plasmon mode to an external electromagnetic field
which is used to probe the system.

The modulating grating on top
of a conducting sheet generates and mixes Bragg modes of a
reflected/transmitted electromagnetic field.
 The existence of the Bloch-like modes due to the modulation
is a direct consequence of the nonlocal mixing of specular and
diffraction modes of the reflected electromagnetic field by
free-electron induced optical polarization.  Furthermore,  there
could be  interference between  a
pair of surface optical-polarization waves with different Bragg order
numbers in the presence of a modulation. The interference of these two
counter-propagating surface waves leads to the formation of a
Wannier-like state with associated electromagnetic fields localized
within the gaps of the modulating potential. These are the effects which contribute
to the physical difference between the induced potential with and
without a periodic potential.

\begin{acknowledgments}
This research was supported by  contract \# FA 9453-07-C-0207 of AFRL.
\end{acknowledgments}

\newpage

\begin{figure}[p]
\epsfig{file=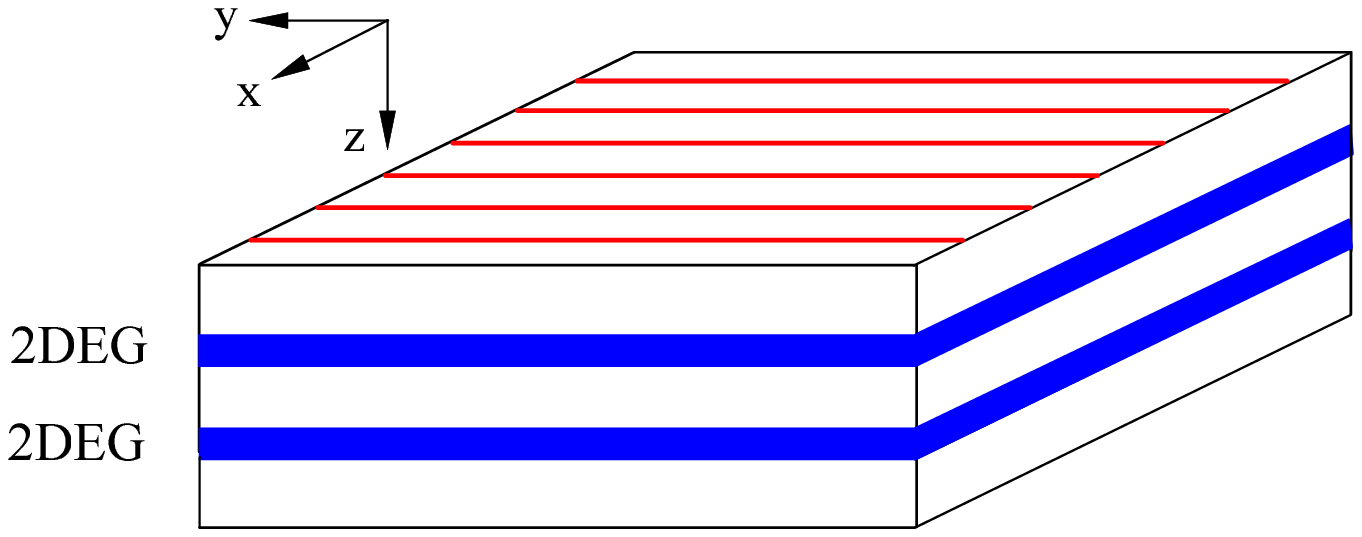,width=5in,height=2.8in}
\caption{Schematic illustration of a bilayer with an applied electrostatic modulation
periodic in the $x$ direction. The surface at $z=0$ is covered by a grating (red lines) with period $d$ and the width of the  grating is assumed
small compared to the period $d$. The doped barriers (AlGaAs)  are the unshaded areas
and quantum wells (GaAs, blue areas) in the $z$ direction are indicated as the 2DEG layers.
 The system is embedded in a dielectric medium.}
\label{FIG:7}
\end{figure}

\begin{figure}[p]
\epsfig{file=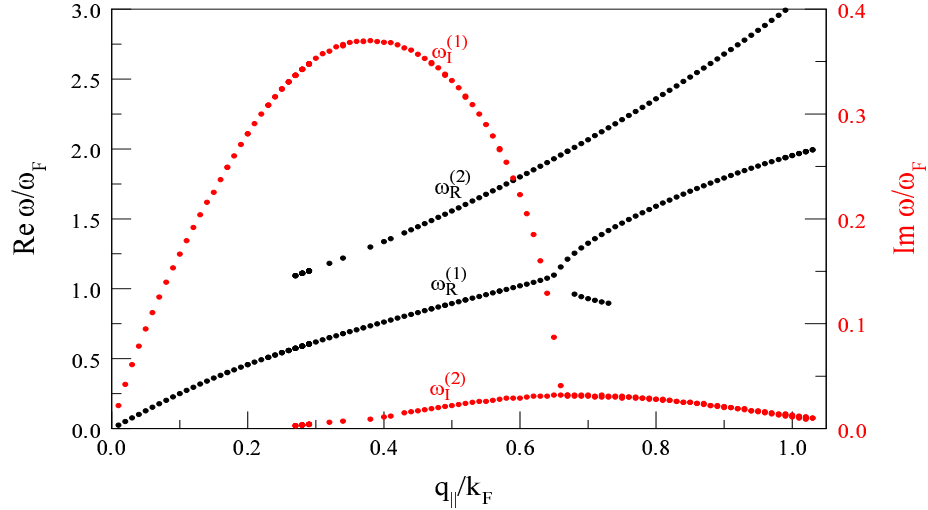,width=5in,height=2.8in}
\caption{Plots of the real (left scale) and imaginary (right scale) parts
of the plasmon frequency for a bilayer 2DEG system with spacing
$a=100$ {\AA} between the layers. The frequencies are expressed
in units of $\omega_F$ and are plotted as functions of the in-plane
wave vector $q_{\parallel}$ in units of $k_F$. The real solutions, shown in black
and labeled by $\omega_R^{(1)}$ and  $\omega_R^{(2)}$,
have  branches which bifurcate. The imaginary solutions,
labeled correspondingly by $\omega_I^{(1)}$ and  $\omega_I^{(2)}$ are shown in red. No
modulating potential was applied in these calculations.
Only those plasmon frequencies with nonzero imaginary parts are presented.}
\end{figure}

\begin{figure}[p]
\epsfig{file=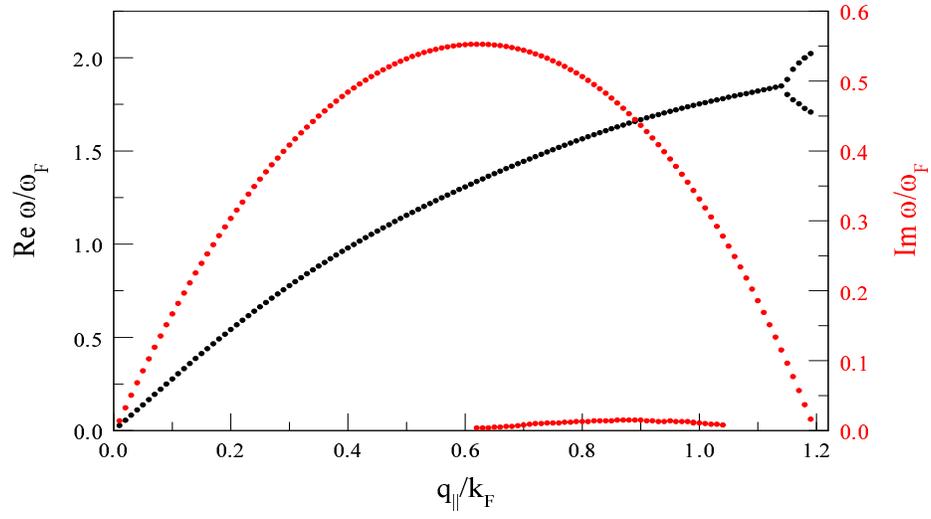,width=5in,height=2.8in}
\caption{The same as Fig.~2, except that  $\omega_p=1.84\omega_F$.}
\end{figure}

\begin{figure}[p]
\epsfig{file=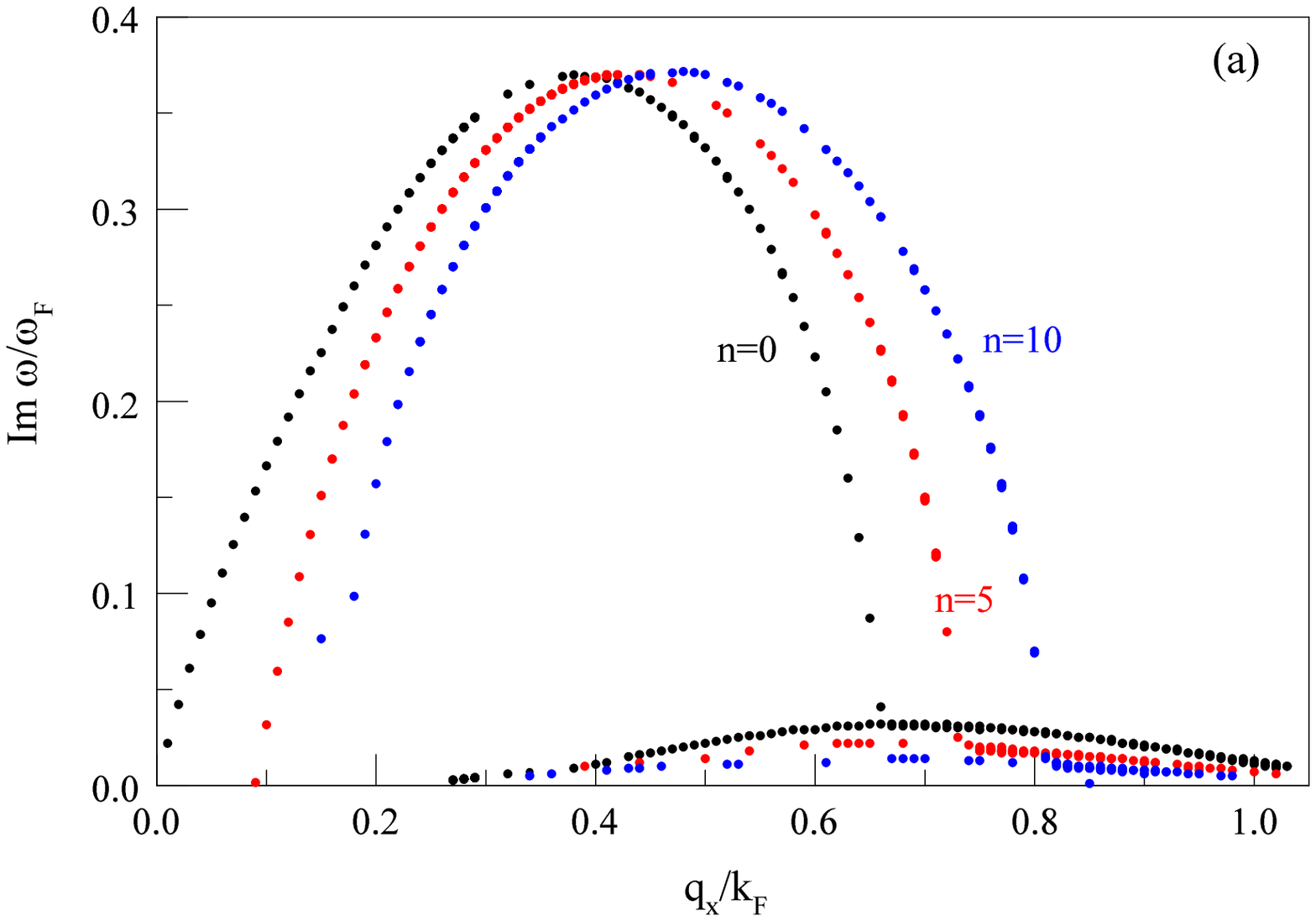,width=5in,height=2.8in}
\epsfig{file=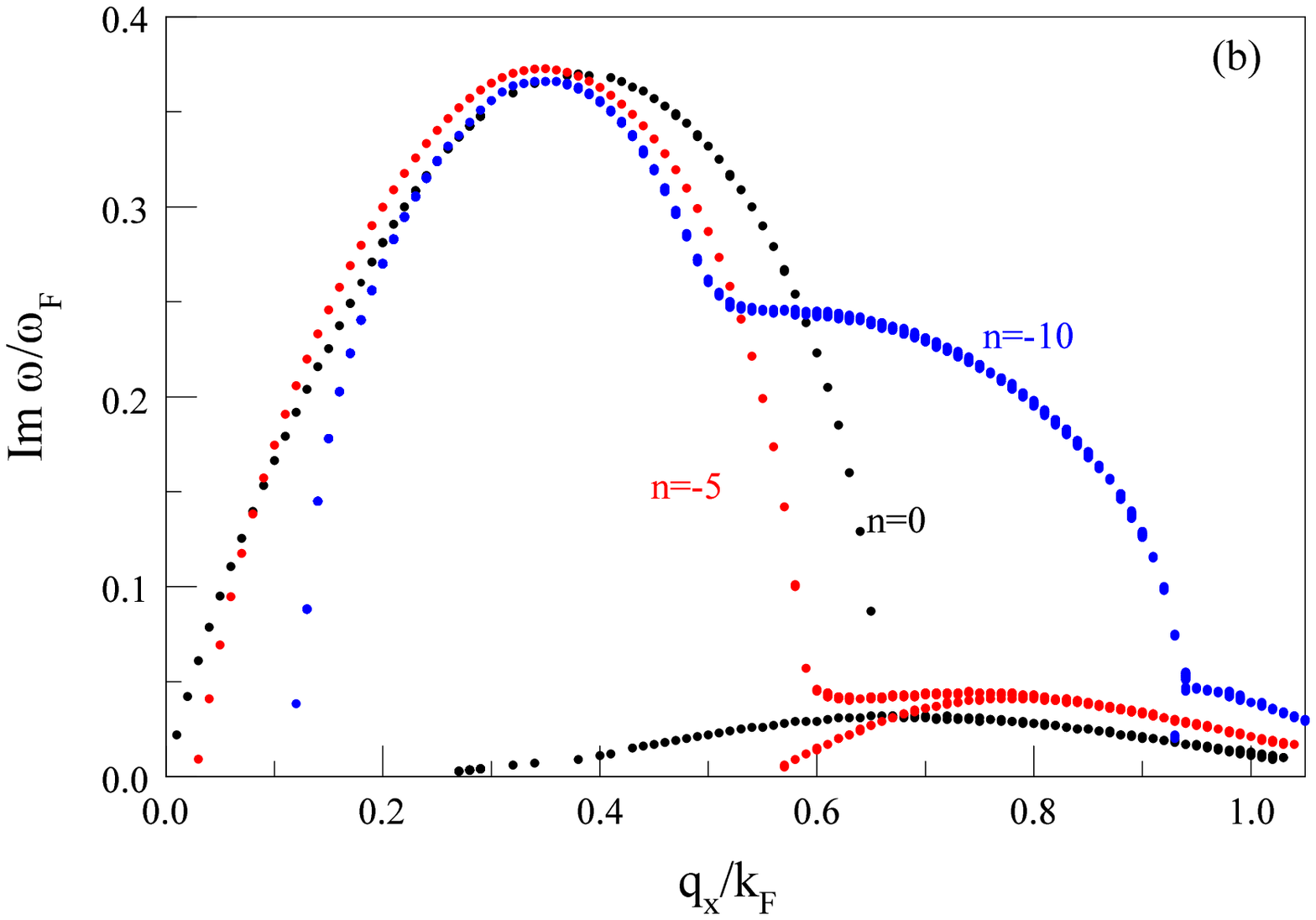,width=5in,height=2.8in}
\caption{(a) Imaginary part of the plasmon frequency in units
of $\omega_F$ for $n=10$ (blue dots) has a much larger $\Im m\; \omega$
than the curves for $n=0$ (black dots) and $n=5$ (red dots).
The parameters used in the calculations are the same as Fig.~2.
The higher-frequency curves are for the acoustic-like plasmons discussed in
Fig.~2. The group of curves with the lower frequency
correspond to the optical plasmons. (b) The same as (a), except that
$n=0,-5,-10$.}
\end{figure}

\begin{figure}[p]
\epsfig{file=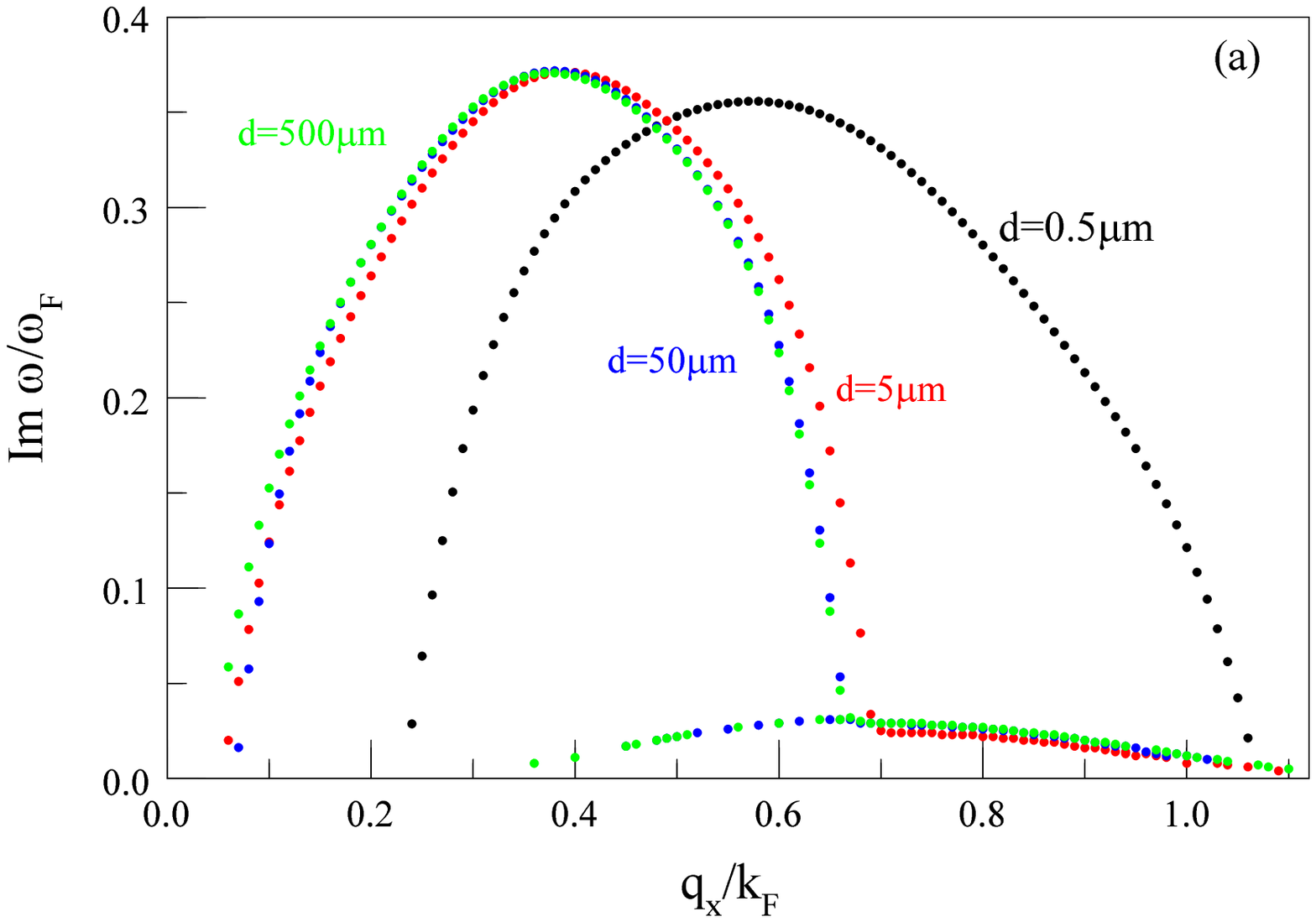,width=5in,height=2.8in}
\epsfig{file=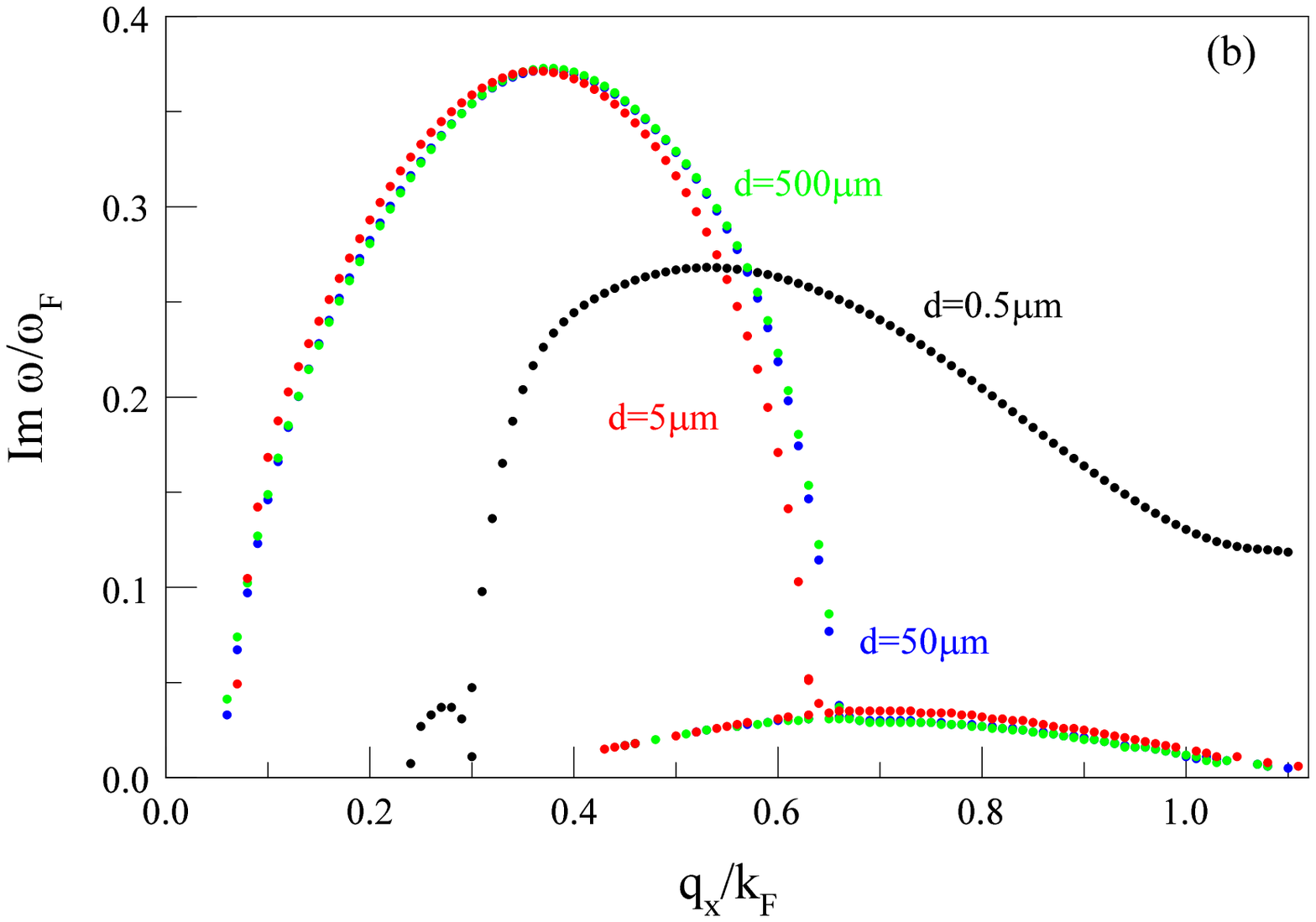,width=5in,height=2.8in}
\caption{The imaginary part of the plasmon frequency
$\omega$ as a function of $q_x$ with $q_y=0$. We  vary the period $d$
of the modulation. Here, $d=0.5\ \mu$m (black dots), $d=50\ \mu$m (red dots)
and $d=500\ \mu$m (green dots).
In  (a),   $n=2$ and (b) $n=-2$. All other parameters
are the same as Fig.~2.}
\end{figure}

\end{document}